\documentclass[iop]{emulateapj}
\usepackage{amsmath}

\begin{document}

\newcommand{\myemail}{jsohn@cfa.harvard.edu}
\newcommand{\kms}{\rm km~s^{-1}}
\newcommand{\kmsmpc}{\rm km~s^{-1}~Mpc^{-1}}
\newcommand{\dn}{D_{n}4000}
\newcounter{mytempeqncnt}

\title{Velocity Dispersions of Brightest Cluster Galaxies and Their Host Clusters}

\author{Jubee Sohn$^{1}$, 
             Margaret J. Geller$^{1}$, 
			Antonaldo Diaferio$^{2,3}$,
			Kenneth J. Rines$^{4}$}

\affil{$^{1}$ Smithsonian Astrophysical Observatory, 60 Garden Street, Cambridge, MA 02138, USA}
\affil{$^{2}$ Universit{\`a} di Torino, Dipartimento di Fisica, Torino, Italy}
\affil{$^{3}$ Istituto Nazionale di Fisica Nucleare (INFN), Sezione di Torino, Torino, Italy}
\affil{$^{4}$ Department of Physics and Astronomy, Western Washington University, Bellingham, WA 98225, USA}

\begin{abstract}
We explore connections between brightest cluster galaxies (BCGs) and their host clusters.
We first construct a HeCS-omnibus cluster sample 
 including 227 galaxy clusters within $0.02 < z < 0.30$;
 the total number of spectroscopic members from MMT/Hectospec and SDSS observations is 52325.
Taking advantage of the large spectroscopic sample, 
 we compute physical properties of the clusters including the dynamical mass and cluster velocity dispersion ($\sigma_{cl}$). 
We also measure the central stellar velocity dispersion of the BCGs ($\sigma_{*, BCG}s$)
 to examine the relation between BCG velocity dispersion and cluster velocity dispersion for the first time. 
The observed relation between BCG velocity dispersion and the cluster velocity dispersion is remarkably tight. 
Interestingly, the $\sigma_{*, BCG} / \sigma_{cl}$ ratio decreases as a function of $\sigma_{cl}$ 
 unlike the prediction from the numerical simulation of \citet{Dolag10}. 
The trend in $\sigma_{*, BCG} / \sigma_{cl}$ suggests that 
 the BCG formation is more efficient in lower mass halos. 
\end{abstract}
                                                                                                                                                                                                                                                                                                                                                                                                                                                                                                                                                                                                                                                                                                                                                                                                                                                                                                                                                                                                                                                                                                                                                                                                                                                                         \section{INTRODUCTION}

Brightest cluster galaxies (BCGs) are a distinctive population of 
 luminous galaxies located in the central regions of galaxy clusters (and groups). 
Identification of the BCGs has a long history since 1784. 
Charles Messier identified a concentration of nebul{\ae} in the Virgo constellation 
 and found its brightest component, M87 \citep{Biviano00}. 
Many studies used BCGs 
 as a tracer for the systematic identification of galaxy clusters 
 in photometric data (e.g. \citealp{Abell58, Abell89, Koester07, Hao10}). 

BCGs are also a unique population that connects galaxy evolution and structure formation models. 
Standard structure formation models predict 
 the hierarchical growth of massive halos of clusters through the stochastic accretion of less massive halos
 \citep{vandenBosch02, McBride09, Zhao09, Fakhouri10, Gao12, Kravtsov12, Haines18}. 
During the mass assembly of the halos, 
 the BCG residing in the bottom of the cluster potential may experience more mergers 
 and thus the evolutionary path of the BCGs differs from the path for less massive galaxies (e.g. \citep{deLucia07}). 
Detailed investigation of BCG properties and the related evolutionary processes
 probe cluster formation models. 

The ratio between stellar mass of a BCG and its halo mass constraints
 both structure formation and galaxy evolution models \citep{Wang06, Moster10, Leauthaud12, Kravtsov18}.
Abundance matching techniques show that 
 this ratio has a peak at $M_{halo} \sim 10^{12} M_{\odot}$ and 
 decreases at higher and lower halo mass (e.g. \citep{Conroy06, Behroozi10, Moster10, Guo10}). 
The decline of the stellar mass fraction at high halo mass suggests that 
 strong feedback processes (including AGN feedback) suppress the stellar mass growth of the BCGs
 \citep{Silk98, McNamara07, Kravtsov12}.  
 
Despite the importance of BCGs, the study of BCGs is not straightforward. 
First of all, the identification of the BCGs is not trivial. 
The Coma cluster is a striking example. 
Coma has two bright galaxies (NGC 4874 and NGC 4889) with a small magnitude difference. 
Neither of these galaxies is located on the dynamical center of the cluster \citep{Rines16}. 
In many cases, 
 an apparently brightest galaxy near the cluster canter has significantly large velocity offset
 with respect to the mean redshift of cluster members ($> 300~\kms$, \citealp{Coziol09, Lauer14, Sohn19c})
 suggesting that these objects may not reside at the bottom of the cluster potential well. 
To avoid confusion, 
 BCG identification requires multi-dimensional data for the galaxies in the cluster field 
 including spatial and radial velocity distribution of the cluster members. 
 
Measuring the physical properties of the BCGs is also not straightforward. 
BCGs usually have very extended stellar halo (cD galaxies, \citealp{Morgan58, Matthews64}) 
 and their profiles overlap the intracluster light. 
Additionally, the BCGs are surrounded by many satellites galaxies. 
The extended stellar halo and contamination from surrounding galaxies affects BCG photometry
 \citep{Gonzalez07, Lauer14} 
\citet{Bernardi13} demonstrate that 
 the photometry of the bright galaxies from the Sloan Digital Sky Survey (SDSS) 
 is systematically offset from the total magnitudes estimated from more complex
 two-dimensional parametric fitting models. 
As \citet{Bernardi13} emphasize, the systematic uncertainty in photometry propagates to the stellar mass 
 and ultimately to the stellar mass to halo mass relation. 
 
Here, we examine the properties of BCGs and their relation with the host cluster properties 
 based on a large spectroscopic sample of galaxy clusters. 
The large spectroscopic sample enables identification of BCGs based on large sets of spectroscopically identified cluster members in multi-dimensional space.
The central stellar velocity dispersion of the BCG itself is insensitive to the galaxy photometry
 \citep{Wake12, Zahid16, Zahid18}. 
We investigate the relationship between stellar velocity dispersion of the BCG and the global cluster velocity dispersion. 
We demonstrate that this relation provides an interesting constraint on structure formation models. 

We describe the cluster and galaxy samples in Section \ref{sample}. 
In Section \ref{catalog}, 
 we introduce the HeCS-omnibus cluster catalog, a large compilation of galaxy clusters with substantial spectroscopic data. 
We also describe the identification of spectroscopic members and the brightest cluster galaxies.  
The HeCS-omnibus cluster catalog includes 227 clusters and the typical number of spectroscopic members per cluster is $\sim180$. 
We explore the connection between the physical properties of BCGs and those of their host clusters in Section \ref{connection}. 
We compare the observed properties of BCGs and their clusters in Section \ref{comparison}. 
We conclude in Section \ref{conclusion}. 
We assume the standard $\Lambda$CDM model with $H_{0} = 70~\kmsmpc$, $\Omega_{m} = 0.3$, $\Omega_{\Lambda} = 0.7$, 
 and $\Omega_{k} = 0.0$ throughout.

\section{SAMPLE}\label{sample}
 
\subsection{Cluster Sample}\label{clsample}

Our goal is to explore the relation between the physical properties of galaxy clusters and 
 the properties of their brightest cluster galaxies (BCGs). 
Spectroscopic surveys yield robust membership identification critical to BCG identification. 
The set of spectroscopically identified members provides 
 physical properties of the galaxy clusters including galaxy velocity dispersion and a basis for deriving the cluster halo mass. 
 
We built a sample of galaxy clusters with substantial spectroscopic data 
 to examine the relation between the BCG and cluster properties.
We first collected data from various spectroscopic surveys.  
The Cluster Infall Region Surveys (CIRS, \citealp{Rines06}) includes 74 nearby clusters with redshift $z < 0.10$. 
\citet{Rines06} collected spectroscopic data from the Sloan Digital Sky Survey (SDSS) Data Release 4 (DR4) 
 and investigated the infall patterns of these clusters. 
We include 71 CIRS clusters with $0.02 < z < 0.10$ in our catalog 
 and compile additional spectroscopic data including SDSS DR14 (see Section \ref{glsample}). 

The Hectospec Cluster Survey (HeCS, \citealp{Rines13, Rines16, Rines18, Rines20}) is
 a large spectroscopic survey of galaxy clusters 
 using the 300 fiber Hectospec mounted on 6.5m Multi-Mirror Telescope (MMT, \citealp{Fabricant05}). 
The first HeCS catalog \citep{Rines13} lists 58 X-ray flux selected clusters with $0.1 < z < 0.3$. 
HeCS-SZ \citep{Rines16} extends the sample by including 123 clusters selected based on Sunyaev-Zel'dovich measurements. 
After removing overlaps with CIRS and HeCS, 
 there are 50 unique clusters added from the HeCS-SZ catalog. 
HeCS-red \citep{Rines18} includes another
 27 high-richness ($\lambda > 64$) redMaPPer clusters with $0.08 < z < 0.29$; 
 we added 23 unique HeCS-red clusters to our sample. 
HeCS-faint is a Hectospec survey of 16 clusters with low X-ray luminosity ($L_{X} < 5 \times 10^{43}$ erg s$^{-1}$, \citealp{Rines20}). 
We exclude 4 HeCS-faint systems not covered by SDSS DR14. 
We include the remaining 12 HeCS-faint systems with $0.04 < z < 0.17$. 
For these HeCS clusters, 
 we compile SDSS DR14 spectroscopy and extensive Hectospec survey data for fainter objects.  
The entire resulting sample includes $\sim 400-550$ redshifts per clusters. 

We also include clusters from ACReS (Arizona Cluster Redshift Survey, \citealp{Haines13}). 
ACReS is a spectroscopic survey also using MMT/Hectospec for 31 clusters from Local Cluster Substructure Survey (LoCuSS). 
ACReS adds 8 additional clusters to our catalog; 4 clusters are not covered by SDSS DR14 and 17 clusters overlap with CIRS or HeCSs. 
Finally, we include three clusters surveyed in other independent Hectospec observational campaigns:
 A68, A611, A1703, A2537 (P.I.: M. Geller)  and A2457 (P.I. : J.Sohn).

Table \ref{clcat} summarizes the number of clusters in the various subsamples we include in the catalog. 
There are a total of 227 clusters in the redshift range $0.02 < z < 0.3$. 
These contain a total of 52325 cluster members. 
Hereafter, we refer to this cluster sample as HeCS-omnibus. 

\begin{deluxetable}{lcc}
\tablecolumns{3}
\tabletypesize{\footnotesize}
\setlength{\tabcolsep}{0.05in}
\tablecaption{The Origin of the HeCS-omnibus Sample}
\tablehead{\colhead{Survey} & \colhead{N\tablenotemark{a}} & \colhead{z range}}
\startdata
CIRS						& 71 & $0.02 < z < 0.10$ \\
HeCS					& 58 & $0.10 < z < 0.29$ \\
HeCS-SZ				& 50 & $0.02 < z < 0.20$ \\
HeCS-red				& 23 & $0.10 < z < 0.26$ \\
HeCS-faint			& 12 & $0.04 < z < 0.17$ \\
ACReS					&	8	& $0.16 < z < 0.29$ \\
Hectospec survey	&	5	& $0.05 < z < 0.28$ \\
\hline
HeCS-omnibus  & 227 & $0.02 < z < 0.29$
\enddata
\label{clcat}
\tablenotetext{a}{The number of unique clusters we add to the HeCS-omnibus sample. }
\end{deluxetable}

\subsection{Galaxy Sample}\label{glsample}

\subsubsection{Photometry}

We use the SDSS Data Release 14 (DR14) galaxy catalog as a basic photometric catalog. 
For individual clusters, 
 we select extended sources brighter than $r_{petro, 0} = 23$ mag 
 within $3^\circ$ of each cluster center.
We use $ugriz$ composite Model (cModel) magnitudes, 
 a linear combination of de Vaucouleurs and model magnitudes. 
We adopt the SDSS foreground extinction for each photometric band. 
Hereafter the photometry refers to extinction-corrected cModel magnitudes. 

\subsubsection{Spectroscopy}

We compiled spectroscopic data for HeCS-omnibus clusters from various surveys. 
We first collect the SDSS DR14 spectroscopy for galaxies with $r < 17.77$. 
The SDSS DR14 spectroscopy significantly improves 
 the spectroscopic sampling of CIRS, originally based on SDSS DR4 spectroscopy. 
SDSS spectra cover $\sim 3800 - 9200$ \AA~ with a spectral resolution of R$\sim 2000$. 
The typical uncertainty in SDSS redshifts is $\sim 7~\kms$. 
Additionally, 
 we collect redshifts from the literature (see the details in \citet{Hwang10})
 through the NASA/IPAC Extragalactic Database (NED). 

The HeCS clusters are extensively surveyed with the MMT/Hectospec. 
We collected the Hectospec spectra through the MMT archive \footnote{http://oirsa.cfa.harvard.edu/archive/search/}.
These Hectospec spectra were acquired through 1.5\arcsec~ radius fibers with a 270 mm$^{-1}$ Hectospec gratings. 
The Hectospec spectrum covers 3700 - 9000~\AA~ with a typical spectral resolution of R$\sim 1700$.

We reduce these spectra homogeneously using HSRed v2.0, an IDL pipeline for reducing the Hectospec spectra. 
We use RVSAO \citep{Kurtz98} to cross-correlate the observed spectra with a set of template spectra
 to measure the redshifts.
RVSAO yields a cross-correlation score $R_{XC}$. 
Following previous HeCS surveys, we use reliable redshifts with a score $R_{XC} > 3$. 
 
The Massive Cluster Survey with Hectospec (MACH)
 is another extended spectroscopic survey using MMT/Hectospec
 for seven massive clusters within $0.06 < z < 0.09$ selected from CIRS \citep{Sohn20}.
MACH is a remarkable spectroscopic campaign for nearby clusters 
 that provides more than 2500 spectra per cluster. 
For example, Abell 2029, used for a pilot study of MACH,
 is one of the best sampled clusters with $\sim 1200$ spectroscopically identified members \citep{Sohn17, Sohn19b}. 
A more detailed discussion of the entire MACH sample will be included in \citet{Sohn20}. 

We also collected the Hectospec spectra for ACReS clusters through the ACReS database\footnote{http://herschel.as.arizona.edu/acres/data/acres\_data.php}. 
ACReS provides redshift measurements based on $\chi^{2}$ minimization.
They also include visual inspection flags; 
 flag 0 means an insecure redshift, flag 1 indicates a less certain redshift, and flag 2 means a secure redshift.  
We use only redshift measurements with visual inspection flag 1 and 2. 

There are six HeCS-omnibus clusters from the OmegaWINGS catalog \citep{Gullieuszik15, Moretti17}: A85, A168, A193, A957, A2399, A2457. 
We collected redshifts from the OmegaWINGS catalog to increase the number of spectroscopic redshifts.  
The OmegaWINGS spectra are obtained using the AAOmega spectrograph 
 and cover $3800 - 9200$ \AA~ with a spectral resolution of $R = 1300$. 
The typical uncertainty of OmegaWINGS redshifts is $\sim 50~\kms$. 
We match the OmegaWINGS redshift catalog with the galaxy catalogs for six clusters
 and update the redshift compilation.

\subsubsection{Stellar Mass}

We derive stellar masses based on the SDSS photometry 
 using the Le Phare fitting code \citep{Arnouts99, Ilbert06}. 
We follow the stellar mass estimation process described in \citet{Sohn17}
 who measured the stellar mass function of spectroscopic members of the nearby massive cluster A2029. 
We briefly review the stellar mass estimation here (see \citealp{Sohn17, Sohn19b}). 

Le Phare computes a mass-to-light ratio 
 by comparing synthetic spectral energy distribution models and SDSS photometry for a galaxy. 
The set of SED models are based on the \citet{BC03} code with the \citet{Chabrier03} initial mass function. 
We run the models with three metallicities, 
 an exponentially declining star-formation with e-folding timescales $\tau = 0.1, 0.3, 1, 2, 3,5, 10, 15, 30$, 
 and stellar population ages between 0.01 and 13 Gyr.
To take foreground extinction into account, 
 we use the \citet{Calzetti00} extinction law with an E(B-V) range of 0.0 to 0.6. 
Based on these SED models, 
 we calculate the probability density function (PDF) for the stellar mass.
We use the stellar mass that is the median of the appropriate PDF. 
  
\subsubsection{$\dn$}

We measure the $\dn$ index, 
 a powerful spectroscopic indicator of the stellar population age of a galaxy (e.g. \citealp{Kauffmann03a}). 
We use the definition given in \citet{Balogh99}; 
 $\dn$ is a ratio between the flux within $4000 - 4100$ \AA~ and the flux within $3850 - 3950$ \AA. 
We estimate the $\dn$ from the cluster galaxy spectra obtained with both SDSS and Hectospec spectrographs. 
Because the $\dn$ values measured from Hectospec and SDSS for the same objects are consistent within $\sim 5\%$ \citep{Zahid17}, 
 we do not apply any additional correction to the $\dn$ measurements. 

\subsubsection{Stellar velocity dispersion}

The stellar velocity dispersions we use are from either SDSS or Hectospec spectra. 
For objects with SDSS spectra, 
 we compile the velocity dispersion measurements from the Portsmouth reduction \citep{Thomas13}. 
\citet{Fabricant13} demonstrated that the Portsmouth velocity dispersions show a tight one-to-one relation with the Hectospec velocity dispersion. 
Thus, we use the Portsmouth and Hectospec velocity dispersions interchangeably without correction. 
The Portsmouth velocity dispersions are measured with the Penalized Pixel-Fitting (pPXF) code \citep{Cappellari04}
 and stellar population templates from \citet{Maraston11}. 
\citet{Thomas13} discuss the details of these velocity dispersion measurements. 

We measure the stellar velocity dispersion from Hectospec spectra
 using the University of Lyon Spectroscopic analysis Software (ULySS, \citealp{Koleva09}). 
We prepare the stellar population templates using the PEGASE-HR code and the MILES stellar library. 
We convolve these templates to the Hectospec resolution with various velocity dispersions. 
Then, Ulyss derives the velocity dispersion based on a $\chi^{2}$ fit of the Hectospec spectra and the templates. 
We limit the fitting range to the rest-frame spectral range $4100 - 5500$ \AA~ to minimize the velocity dispersion uncertainty. 

We apply an aperture correction to derive consistent velocity dispersions from SDSS/Portsmouth and Hectospec data. 
\citet{Zahid16} define the aperture correction: 
 $\sigma_{A} / \sigma_{B} = (R_{A} / R_{B})^{\beta}$. 
Based on 270 objects with both SDSS and Hectospec spectra, 
 \citet{Sohn19b} derived an aperture correction coefficient $\beta = -0.059 \pm 0.014$
 where $R_{A} = R_{SDSS} = 1.5\arcsec$ and $R_{B} = R_{Hecto} = 0.75\arcsec$
We use this coefficient to put the central stellar velocity dispersions on a single system. 

Because HeCS-omnibus clusters are distributed over a wide redshift range, 
 the SDSS and Hectospec fibers cover different physical scale within the BCGs depending on the cluster redshift. 
Thus, we convert the measured stellar velocity dispersion to a fiducial 3 kpc aperture. 
Hereafter, the central stellar velocity dispersion indicates the aperture corrected velocity dispersion within a 3 kpc aperture. 
\citet{Newman13} showed that the velocity dispersion profile of BCGs derived from longslit observations changes little in various BCGs
 for radius $\lesssim 20$ kpc. 
Therefore, the choice of this fiducial radius does not impact the results. 

\section{The HeCS-omnibus Catalog}\label{catalog}

The HeCS-omnibus catalog includes 227 clusters with $0.01 < z < 0.29$. 
We derive the properties of HeCS-omnibus clusters based on this catalog. 
In Section \ref{member}, 
 we describe the cluster membership determination based on spectroscopic redshifts and the caustic technique \citep{Diaferio97, Serra13}. 
We derive the cluster velocity dispersion ($\sigma_{cl}$), and the  characteristic radius and mass ($R_{200}$ and $M_{200}$) in Section \ref{clphysical}. 
Finally, we identify the BCGs in each cluster based on multi-dimensional information 
 including the spatial, color, magnitude, and redshift distributions of cluster members and the BCG (Section \ref{bcgiden}).

\subsection{Membership Determination}\label{member}

The caustic technique \citep{Diaferio97, Diaferio99, Serra13} 
 is a widely used tool for identifying spectroscopic members of a galaxy cluster. 
The caustic technique measures the mass in the infall region of a cluster. 
The technique calculates the escape velocity profile and the corresponding mass profile 
 as a function of clustercentric distance. 
Based on the escape velocity profile in redshift space, 
 the technique identifies spectroscopic members within the trumpet-like caustic pattern \citep{Serra13}. 

Tests based on N-body simulations suggests that 
 the technique identifies $\sim 90\%$ of the true spectroscopic members within $R_{cl} < 3R_{200}$
 when a cluster is well sampled ($N_{\rm member} > 50$, \citealp{Serra13}). 
Furthermore, 
 the technique successfully separates interlopers in the simulations;
 fewer than $8\%$ of caustic members are interlopers. 
The caustic technique has been applied to many large spectroscopic surveys of clusters for member identification 
 (e.g. \citealp{Rines06, Rines13, Rines16, Rines18, Hwang12, Haines13, Habas18, Sohn17, Sohn19b}). 

We identify spectroscopic members of HeCS-omnibus clusters using the caustic technique. 
The HeCS-omnibus clusters have $16 - 1209$ spectroscopic members; the median number is 180. 
These rich samples of spectroscopic members 
 enable detailed analysis of the cluster dynamics \citep{Saro13}. 
Table \ref{cltable} lists the number of spectroscopic members in each cluster.

\subsection{Physical Properties of the HeCS-omnibus Clusters}\label{clphysical}

The caustic technique computes the mass profile of a cluster \citep{Diaferio97}. 
Based on the caustics, 
 we calculate the characteristic mass $M_{200}$ and radius $R_{200}$ of each cluster.
Within $R_{200}$, the mean density is 200 times the critical density at the cluster redshift. 
We also derive the velocity dispersion for the spectroscopic members within $R_{cl} < R_{200}$. 
We use the bi-weight technique \citep{Beers90} to calculate the velocity dispersion. 
We calculate the velocity dispersion uncertainties ($1\sigma$ standard deviation) from 10,000 bootstrap resamplings. 
Table \ref{cltable} lists the $M_{200}$, $R_{200}$, and $\sigma_{cl}$ for each of the HeCS-omnibus clusters. 

We compare the physical properties of HeCS-omnibus clusters 
 with previous values from CIRS and HeCS.
The $M_{200}, R_{200}$ and $\sigma_{cl}$ of the HeCS-omnibus clusters 
 are consistent with the earlier measurements.
Individual measurements can differ randomly by $\sim 20\%$
 largely as a result of the increased sampling here. 
 
\begin{deluxetable*}{lcccccccc}
\tablecolumns{9}
\tabletypesize{\footnotesize}
\setlength{\tabcolsep}{0.05in}
\tablecaption{HeCS-omnibus Clusters}
\tablehead{
\colhead{Cluster ID} & \colhead{R.A.} & \colhead{Decl.} & \colhead{z} & 
\colhead{N$_{\rm mem, caustic}$} & \colhead{N$_{\rm mem, R200}$} & 
\colhead{$\sigma_{cl}$\tablenotemark{a}} & \colhead{$R_{200}$\tablenotemark{b}} & \colhead{$M_{200}$\tablenotemark{b}} \\
\colhead{}				& (J2000)				& (J2000)				& \colhead{}		&
\colhead{}				& \colhead{}		&
\colhead{(km s$^{-1}$)} & \colhead{(Mpc)} & \colhead{$10^14 M_{\odot}$}
}
\startdata
MKW4	& 181.124967	&	1.872426		& 0.0204 &	232	&  102 & $ 473 \pm 47$ & $ 1.12 \pm 0.04$ & $ 1.62 \pm 0.18$ \\
A1367	& 176.175872	&	19.734385	& 0.0225 &	530	&  229 & $ 726 \pm 31$ & $ 1.60 \pm 0.04$ & $ 4.74 \pm 0.41$ \\
MKW11 & 202.361705	&	11.709481	& 0.0233 &	79		&   48 & $ 380 \pm 34$ & $ 0.81 \pm 0.00$ & $ 0.62 \pm 0.01$ \\
A779		& 139.934056	&	33.710087	& 0.0232 &	139	&   48 & $ 296 \pm 34$ & $ 0.72 \pm 0.00$ & $ 0.44 \pm 0.01$ \\
Coma	& 195.000629	&	27.969336	& 0.0234 &	1139 &  672 & $ 873 \pm 21$ & $ 1.76 \pm 0.08$ & $ 6.38 \pm 0.90$
\enddata
\label{cltable}
\tablenotetext{a}{The velocity dispersion measured with the bi-weight technique \citep{Beers90} for galaxies within $R_{200}$. }
\tablenotetext{b}{$R_{200}$ and $M_{200}$ based on the caustic technique. }
\end{deluxetable*}

\begin{figure}
\centering
\includegraphics[scale=0.43]{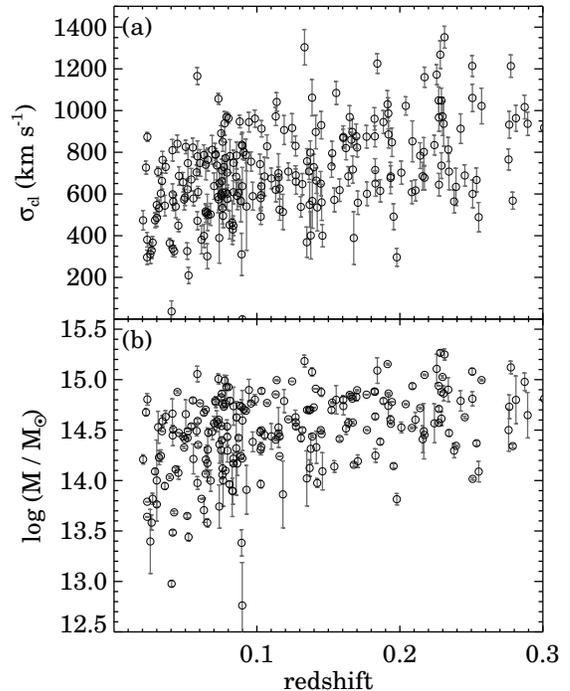}
\caption{(a) The velocity dispersion of HeCS-omnibus clusters as a function of cluster redshift. 
(b) The $M_{200}$ of the clusters based on the caustic technique as a function of cluster redshift. }
\label{clz}
\end{figure}

Figure \ref{clz} shows $\sigma_{cl}$ and $M_{200}$ of the HeCS-omnibus clusters 
 as a function of cluster redshift.
Most HeCS-omnibus clusters have velocity dispersions larger than $400~\kms$
 and dynamical masses $\gtrsim 10^{14} M_{\odot}$. 
Less massive systems are only identified at low redshift ($z \lesssim 0.1$) because of Malmquist bias.

\subsection{Identification of the Brightest Cluster Galaxies}\label{bcgiden}

\subsubsection{Identification of the BCGs}

The BCG is the brightest galaxies in a cluster as the name suggests. 
Conventionally, BCGs are selected based on photometry (e.g. \citealp{Lin04, Lauer14}), 
 although different photometric bands have been used. 
More recently, 
 the galaxy with highest stellar mass among cluster members has been chosen as the BCG 
 (e.g. \citealp{Gozaliasl19}). 
This identification facilitates direct comparison with numerical simulations
 where the central galaxy presumably corresponds to the most massive subhalo in a cluster potential. 
However, identification of BCGs based on stellar mass may introduce uncontrolled systematics
 depending on the color and morphology of the galaxy. 
 
Here, we define the BCG in the $r-$band. 
We additionally required that the BCG candidate be located within  $0.5~R_{200}$. 
We apply this selection to reduce confusion resulting from bright galaxies outside the cluster center. 
Based on these two criteria, we first identify BCG candidates. 

Some BCG candidates are not the actual BCG.
These objects make the first cut due to imperfect SDSS photometry. 
Photometry of galaxies in crowded field is challenging 
 because sky subtraction and masking of other objects are both difficult (e.g. \citealp{Bernardi10}). 
Furthermore, \citet{Bernardi13} show that the SDSS cModel magnitudes we use
 often overestimate the total magnitude of the object. 
They also demonstrated that the deviation is more significant for brighter galaxies in the cluster. 
These photometric issues can confuse BCG identification based on SDSS photometry. 

For some BCGs, more recent SDSS photometry significantly overestimates the magnitudes. 
For example, 
 \citet{Sohn19b} showed that the cModel magnitudes for the BCGs of A2029 and A2033  from SDSS DR12 photometry 
 (similar to the DR14 photometry)
 are $\sim 3-4$ magnitude larger than those from SDSS DR7 photometry. 
The SDSS DR7 photometry is much better agreement with the luminosities of these BCGs in the literature (e.g. HyperLEDA, \citealp{Makarov14}). 
In cases of large disagreement, the genuine BCGs can be misidentified.

We thus visually inspect BCG candidates based on the SDSS images. 
We update the BCG identification if there is an apparently brighter cluster member with incorrect SDSS photometry. 
There are $\sim 25$ HeCS-omnibus clusters with apparent BCGs that are inconsistent with the obvious brightest galaxy. 
Because these visually identified BCGs are bigger, brighter and closer to the cluster center, we also refine the visual identification. 
 
\subsubsection{HeCS-omnibus BCG Catalog}

Table \ref{bcgtable} lists the BCGs of the 220 HeCS-omnibus clusters. 
We cannot identify BCGs for five clusters (A1986, A2537, MS2349+2929, Zw1478, MSPM06300)
 because the obvious BCGs have no spectroscopic redshifts. 
The table includes the SDSS object ID, R.A., Decl., redshift, and $r-$band magnitude of the BCGs. 
We also include the physical properties of the BCGs including $\dn$, stellar mass, and stellar velocity dispersion. 
We note that 171, 216, 180 BCGs have $\dn$, stellar mass, and stellar velocity dispersion measurements, respectively. 

\begin{figure}
\centering
\includegraphics[scale=0.46]{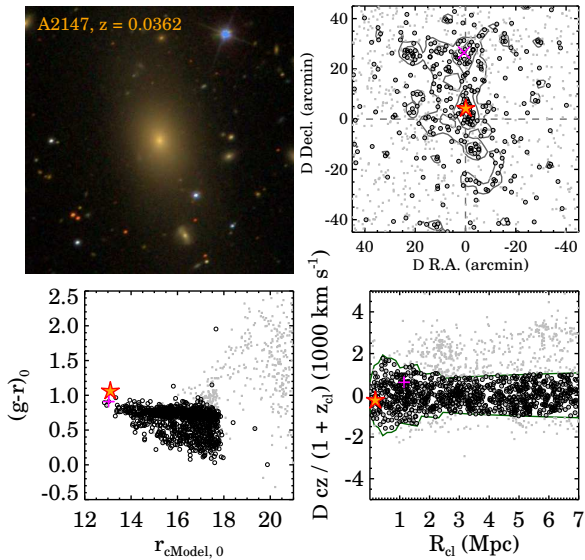}
\caption{Sample plots summarizing BCG identification for A2147.
(Upper left) The SDSS color-composite image of the A2147 BCG (FoV: $20\arcsec \times 20\arcsec$). 
(Upper right) The spatial distribution of galaxies around the cluster center. 
Gray squares are the galaxies with spectroscopic redshifts and black circles are the spectroscopic members. 
The contours show the number density map of spectroscopic members of the cluster.
The red star marks the BCG. 
The magenta cross shows the BCG identified by \citet{Lin04} or \citet{Lauer14}, if any. 
(Lower left) $g-r$ vs. $r$ color magnitude diagram of the cluster field. 
Symbols are the same as in the upper right panel. 
(Lower right) The R-v diagram of the cluster field. 
The solid lines show the caustic pattern for the cluster. }
\label{bcgsummary}
\end{figure}

Figure \ref{bcgsummary} summarizes the BCG identification for HeCS-omnibus clusters:
 (a) the SDSS color-composite image of the BCG, 
 (b) the spatial distribution of cluster members and the BCG with respect to the cluster center, 
 (c) the g-r vs. r color-magnitude diagram, and 
 (d) the R-v diagram of cluster members and the BCG. 
The multi-dimensional graphs confirm that the BCG is indeed the massive central galaxy in each HeCS-omnibus cluster. 
These multi-dimensional graphs for the entire HeCS-omnibus clusters are available in our webpage \footnote{https://www.jubeesohn.com/data}.  
 
\begin{deluxetable*}{lccccccc}
\tablecolumns{8}
\tabletypesize{\footnotesize}
\setlength{\tabcolsep}{0.05in}
\tablecaption{The BCGs of the HeCS-omnibus Clusters}
\tablehead{
\colhead{Cluster ID} & \colhead{BCG Object ID\tablenotemark{a}} & 
\colhead{R.A.} & \colhead{Decl.} & \colhead{z} & 
\colhead{$r_{petro, 0}$} & \colhead{$\log (M_{*}/M_{\odot})$} & \colhead{$\sigma_{*}$}  \\
\colhead{} & \colhead{} & \colhead{(J2000)} & \colhead{(J2000)} & \colhead{} &
\colhead{} & \colhead{} & \colhead{(km s$^{-1}$)}}
\startdata
MKW4	&	1237651735757455411 & 181.112774 &	1.895971		& $ 0.0197 \pm 0.0001$ & $11.78 \pm  0.01$ & $ 11.65 \pm   0.25$ & $ 297 \pm    3$ \\
A1367	&	1237668293912690766 & 176.008980 &	19.949825	& $ 0.0208 \pm 0.0001$ & $12.12 \pm  0.02$ & $ 11.58 \pm   0.18$ & $ 281 \pm    4$ \\
MKW11	&	1237661816564482150 & 202.339834 &	11.735109	& $ 0.0229 \pm 0.0001$ & $13.13 \pm  0.01$ & $ 11.03 \pm   0.19$ & $ -99 \pm  -99$ \\
A779		&	1237661126155436164 & 139.945219 &	33.749742	& $ 0.0230 \pm 0.0001$ & $12.10 \pm  0.01$ & $ 11.53 \pm   0.22$ & $ -99 \pm  -99$ \\
Coma	&	1237667444048723983 & 195.033862 &	27.976941	& $ 0.0215 \pm 0.0001$ & $11.71 \pm  0.01$ & $ 11.47 \pm   0.19$ & $ 368 \pm    5$
\enddata
\label{bcgtable}
\tablenotetext{a}{SDSS DR14 object ID.}
\end{deluxetable*}
 
\subsubsection{Comparison with Previous BCG Catalogs}

We cross-check our BCG identification with previous BCG catalogs. 
\citet{Lin04} publish a catalog of BCGs in 93 clusters and groups. 
There are 29 HeCS-omnibus clusters that overlap with \citet{Lin04} and 27 BCGs correspond exactly to the BCGs identified in \citet{Lin04}. 
We identify different BCGs for two clusters: A2065 and A2147. 
In both cases, the BCGs identified by \citet{Lin04} are not coincident with the cluster center ($R_{cl} > 0.7 R_{200}$). 

The BCG catalog from \citet{Lauer14} includes 49 HeCS-omnibus clusters. 
We select different galaxies as BCGs for five of their clusters: A1066, A1436, A2065, A267, and A602. 
We inspect these difference in BCG identification based on location, magnitude, and morphology of the previously selected BCGs. 
The BCGs we identify are either brighter or they are located closer to the cluster center than the previously identified object. 
We also compare with the BCG catalog from \citet{Kluge19} that includes 170 BCGs. 
Among 42 matched clusters, 
 the BCGs of six clusters (A1066, A1423, A2065, A2199, A602, MKW4) are inconsistent. 
Again, the BCGs we identify are closer to the caustic center than the galaxies identified in \citet{Kluge19}. 

\subsubsection{Properties of the BCGs}

We examine the physical properties of the BCGs including $\dn$, stellar mass, and central velocity dispersion. 
For comparison, we investigate the properties of SDSS field galaxies.
The field comparison sample is from the SDSS spectroscopic galaxy sample with magnitude limit $r < 17.77$ and 
 redshift range $z < 0.1$. 
We obtain $\dn$, stellar mass, and the central velocity dispersions from the Portsmouth reduction. 

\begin{figure}
\centering
\includegraphics[scale=0.46]{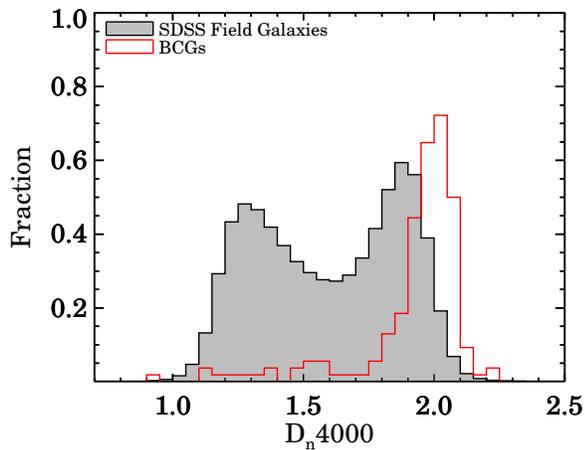}
\caption{The $\dn$ distributions of the HeCS-omnibus BCGs (open histogram) and 
 SDSS field galaxies (filled histogram).}
\label{dn}
\end{figure}

Figure \ref{dn} displays the $\dn$ distributions of the BCGs and the SDSS field galaxies. 
The BCG $\dn$ distribution clearly differs from field. 
The majority of the BCGs ($\sim 95\%$) are quiescent galaxies with $\dn > 1.5$. 
Unlike the BCGs, SDSS field galaxies show an obvious bimodal distribution. 
Furthermore, the field quiescent population shows a peak at $\dn \sim 1.85$, 
 but the BCG population shows a peak at $\dn \sim 2.0$ with a typical uncertainty of 0.03. 
This comparison suggests that 
 the stellar population of the central region of BCG is old. 

\begin{figure}
\centering
\includegraphics[scale=0.46]{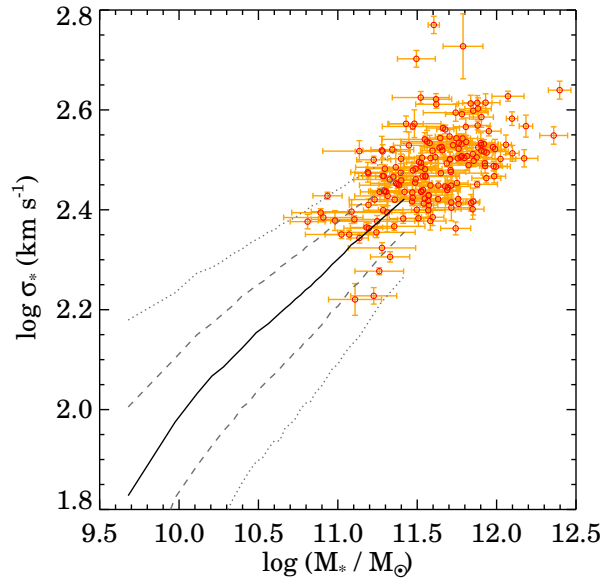}
\caption{Central velocity dispersion vs. stellar mass for the HeCS-omnibus BCGs (red circles). 
The solid line is the same relation for SDSS galaxies. 
The dashed and dotted lines include the 68\% and 95\% for SDSS galaxies. }
\label{vdispmass}
\end{figure}

Figure \ref{vdispmass} shows 
 the central velocity dispersion as a function of stellar mass for BCGs. 
Most of the BCGs have high stellar mass ($> 10^{11.5} M_{\odot}$) and high velocity dispersion ($> 250~\kms$). 
We also plot the quiescent population in the SDSS field sample with $\dn > 1.5$. 
The solid line in Figure \ref{vdispmass} indicates the mean relation for SDSS field quiescent galaxies
 and the dashed (dotted) lines show the boundaries that include 68\% (95\%) of field galaxies.
The comparison clearly shows that the BCGs represent the most massive tail of the population in terms of both stellar mass and velocity dispersion. 
Interestingly, the BCGs do follow the relation between stellar velocity dispersion and stellar mass defined by other quiescent galaxies. 

\section{Connection between BCGs and Clusters}\label{connection}

The HeCS-omnibus clusters provide a basis for examining relations between BCG properties 
 and the dynamical properties of their host clusters. 
We use absolute magnitude, stellar mass, and stellar velocity dispersion 
 as mass proxies for the BCGs. 
We also use the cluster velocity dispersion and caustic mass (dynamical mass) for probing cluster properties. 
Here we explore the relations among these properties. 

\begin{figure}
\centering
\includegraphics[scale=0.4]{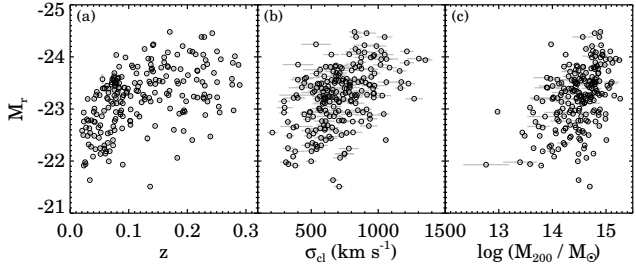}
\caption{$r-$band absolute magnitudes of BCGs as a function of cluster (a) redshift, (b) velocity dispersion, and (c) $M_{200}$. }
\label{lum}
\end{figure}

Figure \ref{lum} (a) shows the absolute magnitude of BCGs in the $r-$band as a function of redshift. 
Overall, BCGs are very bright ($M_{r} < -23$) over the entire range.
Less luminous BCGs are mainly located in low redshift clusters. 
These low redshift clusters are also less massive than higher redshift systems (Figure \ref{clz}). 

Figures \ref{lum} (b) and (c) show the absolute magnitudes of BCGs as a function of $\sigma_{cl}$ and $M_{200}$. 
The clusters with higher velocity dispersion and higher dynamical mass host the brighter BCGs, 
 consistent with previous studies (e.g. \citealp{Lin04}). 
The Spearman rank correlation coefficients for these relations are 0.47 and 0.45
 with significance of $2.66 \times 10^{-13}$ and $2.36 \times 10^{-12}$, respectively.

\begin{figure}
\centering
\includegraphics[scale=0.4]{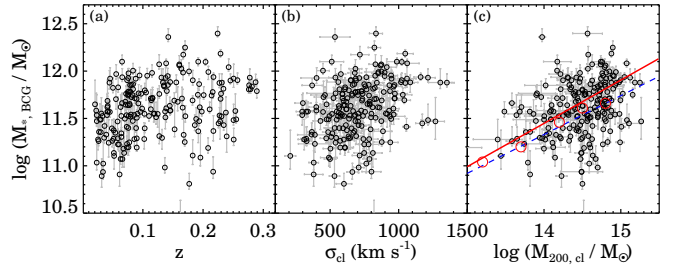}
\caption{Stellar masses of the BCGs as a function of (a) redshift, (b) $\sigma_{cl}$, and (c) $M_{200}$. 
In panel (c), the dashed line shows the BCG stellar mass - cluster halo mass relation from \citet{Erfanianfar19}. 
The solid line displays the best-fit relation we derive for the HeCS-omnibus clusters. }
\label{mass}
\end{figure}

Figure \ref{mass} shows the same relations as Figure \ref{lum}, but based on BCG stellar mass ($M_{*, BCG}$). 
Similar to Figure \ref{lum} (a), 
 the high redshift clusters contain more massive BCGs because our sample includes more massive clusters at higher redshift. 
Figures \ref{mass} (b) and (c) indicate that more massive clusters tend to host more massive BCGs.  
The Spearman correlation coefficients for $M_{*, BCG}$ versus $\sigma_{cl}$ relation and for $M_{*, BCG}$ versus $M_{200}$ relation are
 0.34 and 0.34 with the significance of $3.32 \times 10^{-7}$ and $3.24 \times 10^{-7}$, respectively. 
Note that this correlation is somewhat weaker than the correlation between $M_{r}$ and the global cluster properties. 

Previous observations show similar relations based on various cluster samples covering wide mass and redshift ranges 
 \citep{Kravtsov18, OlivaAltamirano14, Erfanianfar19}. 
Most recently, 
 \citet{Erfanianfar19} investigated the connection between the stellar mass of the BCG and the cluster halo mass 
  based on a large sample of 526 clusters within $0.1 < z < 0.65$. 
They estimated the halo mass of a cluster based on X-ray luminosity and the scaling relation between $M_{200}$ and the X-ray luminosity. 
They derived a best-fit relation between the BCG stellar mass and the cluster halo mass 
 in two samples covering the redshift ranges ($0.1 \leq z \leq 0.3$ and $0.3 < z \leq 0.65$). 
They demonstrated that more massive clusters tend to have more massive BCGs. 
Furthermore, there is no significant redshift dependence for this relation. 
The blue dashed line in Figure \ref{mass} (c) displays the relation from \cite{Erfanianfar19} 
 for clusters in the redshift range ($0.1 \leq z \leq 0.3$) similar to the HeCS-omnibus clusters. 

To compare with previous results, 
 we derive the best-fit relation based on a Markov chain Monte Carlo (MCMC) approach. 
We take the uncertainties in both variables into account and we assume that the uncertainties follow a 2D Gaussian. 
The best-fit relation (red solid line in Figure \ref{mass} (c)) between the BCG stellar mass and the cluster dynamical mass is:
\begin{equation}
\log (M_{*, BCG} / M_{\odot}) = (0.46 \pm 0.03) \log (M_{200, cl} / M_{\odot}) + (5.05 \pm 2.17). 
\end{equation}
The slope of the relation is consistent with the relations in the literature:
 e.g. $0.32 \pm 0.09$ from \citet{OlivaAltamirano14} and $0.41 \pm 0.04$ from \citet{Erfanianfar19}. 
 
\begin{figure*}
\centering
\includegraphics[scale=0.76]{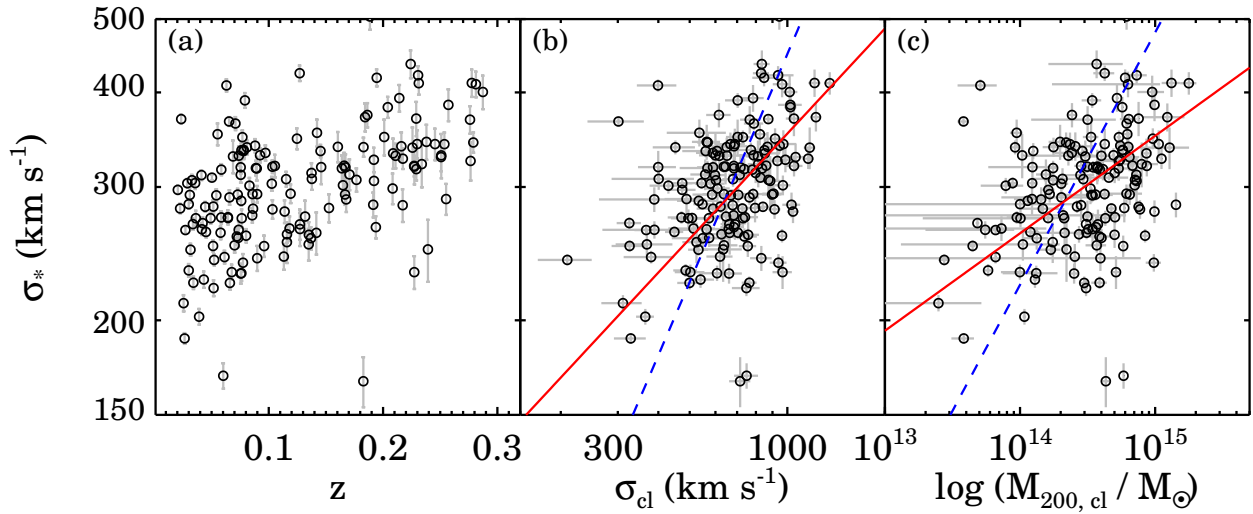}
\caption{Stellar velocity dispersion of BCGs as a function of cluster (a) redshift, (b) $\sigma_{cl}$, and (c) $M_{200}$.
The red solid lines in panel (b) and (c) show the best-fit power law for the BCG stellar velocity dispersion and the cluster halo mass.
The blue dashed lines are the prediction from numerical simulations in \citet{Dolag10} (see Section \ref{comparison}).  }
\label{vdisp}
\end{figure*}

In Figure \ref{vdisp}, 
 we explore the relation between the BCG stellar velocity dispersion ($\sigma_{*, BCG}$)
 and the cluster (a) redshift, (b) $\sigma_{cl}$, and (c) $M_{200}$. 
Interestingly, $\sigma_{*, BCG}$ correlates well with $\sigma_{cl}$ and with $M_{200}$.
These correlations are expected because the stellar velocity dispersion of the central galaxy is a good tracer of its halo mass
 (e.g. \citealp{Wake12, Zahid16, Zahid18}). 

Even though the dynamic range of the BCG stellar velocity dispersion is small,
 the relation between $\sigma_{*, BCG}$ and $\sigma_{cl}$ is remarkably tight (Figure \ref{vdisp} (b)). 
The Spearman rank correlation coefficient is 0.42 with a significance of $3.95 \times 10^{-8}$. 
The best-fit relation (the red solid line) for these variables based on the MCMC approach is 
\begin{equation}
\log \sigma_{*, BCG} = (0.46 \pm 0.09)~ \log \sigma_{cl} + (1.16 \pm 3.45). 
\end{equation}

For the first time, Figure \ref{vdisp} (c) shows that 
 $\sigma_{*, BCG}$ correlates well with the cluster mass based on a large dataset.  
The Spearman rank correlation coefficient for this relation is 0.38 with a significance of $1.27 \times 10^{-6}$. 
The best-fit relation (the red solid line) from the MCMC approach is 
\begin{equation}
\log \sigma_{*, BCG} = (0.13 \pm 0.01)~ \log (M_{200, cl} / M_{\odot}) + (0.61 \pm 3.19). 
\end{equation}

The blue dashed line in Figures \ref{vdisp} (b) and (c) display 
 the expected relations based on cosmological hydrodynamic simulations from \citet{Dolag10}. 
We compare with the \citet{Dolag10} simulations 
 because they are unique in examining the relations between BCG velocity dispersion, cluster velocity dispersion, and halo mass. 
The typical mass resolution of their simulations is $3.1 \times 10^{9} h^{-1}~M_{\odot}$ for dark matter particle and 
 $0.48 \times 10^{9} h^{-1}~M_{\odot}$ for gas particle. 
The simulation uses a smoothed particle hydrodynamics and 
 takes radiative cooling, heating by a UV background, star formation and feedback into account . 
Based on the modified SUBFIND algorithm \citep{Springel01, Dolag09}, 
 they identified 44 clusters with more than 20 satellite galaxies. 
There are star particles not bound to any subhalo within the cluster potential. 
These star particles are either the stellar component of the BCG (cD galaxy) or a diffuse stellar component (DSC). 
\citet{Dolag10} separated these two components 
 based on velocity histograms and derived the velocity dispersion of each of these two components. 
From a Maxwellian fit to the two differerent stellar components, 
 they compute the velocity dispersion of the BCG and the DSC.  

The observed data scatter around the predicted relation from the numerical simulation. 
We note that the BCG velocity dispersion estimates from \citet{Dolag10}, 
 the Maxwellian fit to the BCG stellar component, 
 are not identical to our stellar velocity dispersion estimates. 
\citet{Dolag10} also used the virial mass; we use a proxy for the virial mass, $M_{200}$.
Therefore, differences in the slope of the relations in Figure \ref{vdisp} (c) 
 may result from the different definition of the velocity dispersion and the cluster mass. 
Despite the different slopes,   
 both observation and simulation demonstrate that 
 the stellar velocity dispersion correlates well with the cluster mass. 
Furthermore, the tight relation suggests that the stellar velocity dispersion of the BCGs is a good halo mass predictor 
 in analogy with the stellar mass \citep{Pillepich18}. 

\section{COMPARISON WITH SIMULATIONS}\label{comparison}

Based on 227 HeCS-omnibus clusters, 
 we explore the relation between the BCGs and their host clusters. 
We use three different mass proxies for the BCGs 
 including absolute magnitude, stellar mass, and the stellar velocity dispersion. 
We also use the cluster velocity dispersion ($\sigma_{cl}$) and the cluster dynamical mass ($M_{200}$) measured from the caustic technique 
 to probe the cluster halo mass. 
In general, more massive cluster contain the more massive BCGs. 
Particularly, the BCG stellar velocity dispersion ($\sigma_{*, BCG}$) show a tight relation with cluster halo mass proxies. 
Here, we compare the observed relations with predictions from numerical simulations. 
In Section \ref{simul_mass}, we compare the observed and predicted relations for $M_{*, BCG}$ and $M_{200}$. 
We also investigate the relations between $\sigma_{*, BCG}$ and $\sigma_{cl}$, 
 a distinctive comparison based on the HeCS-omnibus sample (\ref{simul_vdisp}). 

\subsection{$M_{*, BCG}$ vs. $M_{200}$ Relations}\label{simul_mass}

The stellar mass of BCGs in the HeCS-omnibus clusters is correlated with the cluster mass. 
This observed relation is consistent with results from previous observations 
 based on different cluster samples over a wide redshift range.
The relation between $M_{*, BCG}$ and $M_{200}$ provides a testbed for modeling the formation and evolution of galaxy and its halo. 

\begin{figure}
\centering
\includegraphics[scale=0.5]{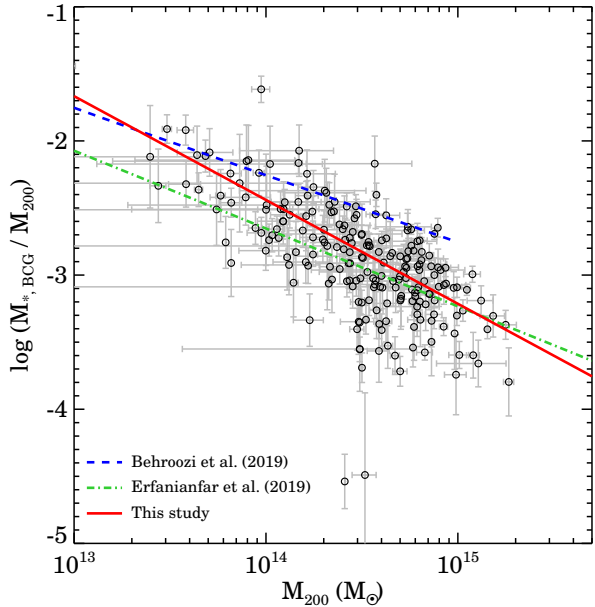}
\caption{
Ratio between the BCG stellar mass ($M_{*, BCG}$) and the cluster dynamical mass ($M_{200}$)
 as a function of cluster dynamical mass ($M_{200}$). 
The red solid line shows the best-fit relation to the HeCS-omnibus sample. 
The green dash-dotted line indicates the best-fit relation for the clusters at $0.1 < z < 0.3$ from \citet{Erfanianfar19}. 
The blue dashed line shows the relation from empirical models of \citet{Behroozi19}. } 
\label{massrat}
\end{figure}

Another important relation to test is the ratio between BCG stellar mass and the cluster halo mass
 as a function of halo mass. 
Figure \ref{massrat} illustrates this relation based on BCG stellar mass ($M_{*, BCG}$) and the cluster dynamical mass ($M_{200}$). 
The observed relation for the HeCS-omnibus clusters shows a tight negative correlation; 
 the Spearman correlation coefficient is -0.68 with a significance of $1.35 \times 10^{-30}$. 
The best-fit relation based on the MCMC approach is: 
\begin{equation}
\log (M_{*, BCG} / M_{200}) = (-0.77 \pm 0.12) \log M_{200} + (8.39 \pm 4.05).
\end{equation}
For comparison, we plot the same relation from \citet{Erfanianfar19} (blue dotted line). 
\citet{Erfanianfar19} estimated the stellar mass based on SDSS, Galaxy Evolution Explorer (GALEX), and
 Wide-Field Infrared Survey Explorer (WISE) photometry using Le Phare. 
They converted cluster X-ray luminosity into the $M_{200}$ based on a scaling relation from \citet{Leauthaud10}. 
Therefore, the slight differences result from the different definitions of $M_{200}$ and the different photometry used for stellar mass estimation. 

We also compare the relation from UNIVERSEMACHINE \citep{Behroozi19}, 
 an empirical model that traces galaxy evolution 
 based on dark matter halo evolution. 
This model is constrained by various observed relations including environmental effects on star formation rate of a galaxy. 
This model provides a stellar mass to halo mass relation for quiescent central galaxies as a function of redshift. 
We compute the relation from UNIVERSEMACHINE (blue dashed line in Figure \ref{massrat}) at the median redshift of HeCS-omnibus clusters, 
 i.e. $z = 0.1$.
The model relation shows a slightly different slope and differs from the observed relation 
 especially at large $M_{200}$ where the data barely overlap with the model.

\subsection{$\sigma_{*, BCG}$ vs. $\sigma_{cl}$ Relations}\label{simul_vdisp}

Based on the extensive dataset for $\sigma_{*, BCG}$, $\sigma_{cl}$, and $M_{200}$ for the HeCS-omnibus clusters, 
 we compare the observed relations among these variables with the same relations from the numerical simulations of \citet{Dolag10}. 
Similar to $M_{*, BCG}$, 
 $\sigma_{*, BCG}$ shows a good correlation with both $\sigma_{cl}$ and $M_{200}$. 
One interesting point is that the measured $\sigma_{*, BCG}$ is generally below $\sigma_{cl}$ as in the simulations. 
\citet{Dolag10} conclude that $\sigma_{*, BCG}$ is governed by the local galactic potential rather than by the global cluster potential. 
The observed relation supports this idea. 

\begin{figure}
\centering
\includegraphics[scale=0.5]{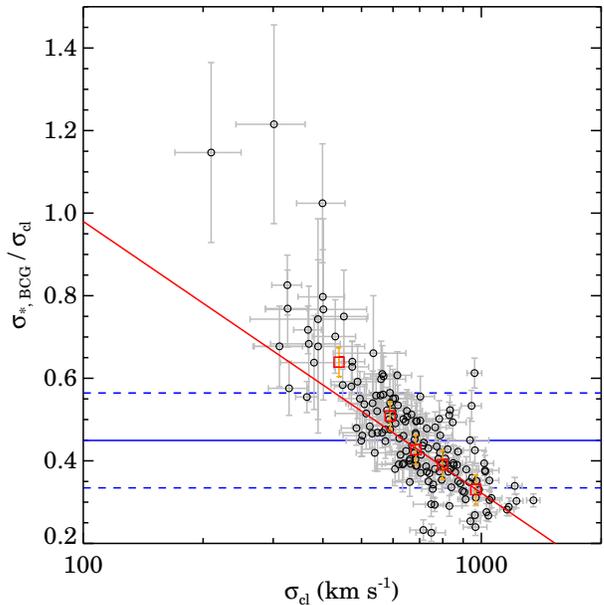}
\caption{
Ratio between the BCG stellar velocity dispersion ($\sigma_{*, BCG}$) and the cluster velocity dispersion ($\sigma_{cl}$) 
 as a function of cluster velocity dispersion ($\sigma_{cl}$). 
Red square are the median $\sigma_{*, BCG} / \sigma_{cl}$ ratio in various $\sigma_{cl}$ bins.
The red dashed line shows the best-fit relation to the HeCS-omnibus sample. 
The blue solid line shows the expected ratio from the cosmological hydrodynamic simulations by \citet{Dolag10}; 
 the blue dashed lines are the $1\sigma$ boundaries for the expected ratio.}
\label{sigrat}
\end{figure}

Figure \ref{sigrat} demonstrates that the ratio between $\sigma_{*, BCG}$ and $\sigma_{cl}$ decreases as a function of $\sigma_{cl}$. 
These two variables show a remarkable tight negative correlation (tighter than the $M_{*, BCG}/M_{200}$ vs. $M_{200}$ relation); 
 the Spearman rank correlation coefficient is -0.79 with a significance of $1.93 \times 10^{-35}$. 
The best-fit relation (the red solid line) based on the MCMC approach is 
\begin{equation}
\sigma_{*, BCG} / \sigma_{cl} = (-0.82 \pm 0.17) \log \sigma_{cl} + (2.77 \pm 3.93).
\end{equation}
This relation indicates that the fraction of mass enclosed in the BCG subhalo continuously decreases as the cluster mass increases. 

Unlike the observed relation, 
 the \citet{Dolag10} simulation suggests a constant ratio between $\sigma_{*, BCG}$ and $\sigma_{cl}$ over a wide range of $\sigma_{cl}$. 
In Figure \ref{sigrat}, the blue solid and dashed lines show the relation and $1\sigma$ deviation from the simulation. 
In this simulation,
 the velocity dispersion of the BCGs, DSC, and cluster galaxies correlate well
 with the virial mass of the cluster halo. 
The cluster halo mass is proportional to $\sigma^{3}$ 
 although the normalization varies with the particular source of the velocity dispersion
 (i.e. $\sigma_{*, BCG}, \sigma_{DSC}$ and $\sigma_{cl}$), 
Thus, the $\sigma_{*, BCG}/\sigma_{cl}$ ratio does not change as a function of $\sigma_{cl}$. 

The observed $\sigma_{*, BCG}/ \sigma_{cl}$ does depend on cluster mass and 
 suggests that the mass fraction associated with 
 the BCGs reflects the evolution of the BCGs and their host halo. 
The high mass clusters are presumably developed systems and their central BCGs have only experienced 
 minor interactions very recently. 
Thus, the center of the BCGs are relaxed and the BCG velocity dispersion has not increased 
 with the cluster velocity dispersion. 
Indeed, \citet{Edwards19} show that the core region of the BCG formed very early ($> 13$ Gyrs ago)
 based on the Integral Field Unit (IFU) observations of BCGs in massive clusters. 
In contrast, galaxies in lower mass clusters encounter one another at relatively low velocities. 
The low relative velocities among members result in a merging instability among member galaxies. 
In these systems, the BCGs can grow more efficiently through more major mergers.

The observed relation between $\sigma_{*, BCG}$ and $\sigma_{cl}$ promises an important test of galaxy and cluster formation models. 
Many previous studies focused instead on the stellar mass of the BCGs. 
Compared with the BCG stellar mass, 
 the BCG velocity dispersion measurement from numerical simulations is insensitive to systematic biases 
 introduced by various baryonic physics inserted in the simulations including feedback models. 
From theoretical point of view the BCG velocity dispersion is also less sensitive to systematics than the stellar mass. 
The BCG velocity dispersion is a strong test of the physics of BCG formation.

\section{CONCLUSION}\label{conclusion}

HeCS-omnibus is a new cluster data compilation including 227 clusters covering the range $0.02 < z < 0.29$
 with $\sim 180$ spectroscopic members per cluster. 
We obtained the spectroscopic survey data mainly from MMT/Hectospec and the SDSS.  
Each of the HeCS-omnibus clusters typically includes $\sim 180$ spectroscopic members. 

We derive physical properties of the cluster galaxies including absolute magnitude, $\dn$, stellar mass, and stellar velocity dispersion. 
We also compute velocity dispersions and dynamical masses of the HeCS-omnibus clusters based on spectroscopic members. 
We identify the BCGs 
 based on multi-dimensional data including the spatial distribution and R-v diagram of spectroscopic members. 

The BCG properties correlate with the mass of the host clusters; 
 more massive clusters tend to have brighter, more massive BCGs. 
These relations for HeCS-omnibus clusters are consistent with previous observations \citep{Kravtsov18, Erfanianfar19}. 
However, we note that the luminosity and the stellar mass of the BCGs suffer from systematic issues 
 due to problematic photometry in crowded region like cluster cores (e.g. \citealp{Bernardi13}). 

The BCG stellar velocity dispersion ($\sigma_{*, BCG}$) show 
 a remarkably tight correlation with host cluster mass proxies ($\sigma_{cl}$ and $M_{200}$). 
This observed relation is consistent with predictions of the numerical simulation of \citet{Dolag10}. 
The tight relation suggests that $\sigma_{cl}$ of a BCG is a good tracer of the cluster halo mass 
 as well as of the BCG stellar mass. 
 
The hierarchical structure formation model predicts 
 a connection between the (stellar) mass of the central galaxy and the cluster halo mass 
 (e.g. \citealp{deLucia07, Pillepich18, RagoneFigueroa18}).
A simple explanation for this connection is that 
 the massive cluster has experienced more accretion (or mergers) and thus its BCG accretes more mass. 
Indeed, many numerical simulations suggest that 
 the mass growth of the BCG is dominated by accretion rather than in situ star formation 
 \citep{deLucia07, RagoneFigueroa18, Pillepich18}. 

Many previous studies have investigated the connection between BCG and cluster halo masses using simulations and observations,
 but most studies are based on the stellar mass.
Our results suggest that the stellar velocity dispersion provide an additional important constraint on this connection.  
Unlike the stellar mass estimates, 
 stellar velocity dispersion is relatively insensitive to systematic issues 
 introduced by photometry in crowded region and or on the assumptions for stellar population models. 
Once the stellar velocity dispersion is measured in simulations
 using the observational procedure we use \citep{Zahid18}, 
 the central velocity dispersions of the BCG
 has the potential to provide additional powerful constraints on formation models. 

We demonstrate that the ratio between the BCG and cluster velocity dispersions, $\sigma_{*, BCG} / \sigma_{cl}$,
 decreases as $\sigma_{cl}$ increases. 
This observed relation suggests that 
 the mass assembly of the BCG subhalo changes in a way that correlates closely with the cluster mass and its accretion history. 
The observed trend is similar to the relation between $M_{*, BCG} / M_{200}$ and $M_{200}$
 shown in previous observational and theoretical works \citep{Kravtsov18, Erfanianfar19, Pillepich18, Behroozi19}. 
Nonetheless, the $\sigma_{*, BCG} / \sigma_{cl}$ ratio trend differs from the theoretical prediction of \citet{Dolag10}
 suggesting that the BCG growth is more efficient in lower mass systems. 
Further tests of $\sigma_{*, BCG}$ and $\sigma_{cl}$ in large-scale numerical simulations
 will be useful for understanding the apparent change in BCG formation efficiency with the cluster mass (and velocity dispersion). 
 
\acknowledgments

We thank Steven Wilhelmy and Zach Schutte for assistance in the early stages of this project. 
J.S. is supported by the CfA Fellowship. 
M.J.G. acknowledges the Smithsonian Institution for support.
AD also acknowledges partial support from the INFN grant InDark and
  from the Italian Ministry of Education, University and Research (MIUR) under the {\it Departments of Excellence} grant L.232/2016.
This paper includes data produced by the OIR Telescope Data Center in the Smithsonian Astrophysical Observatory.
This research has made use of NASA’s Astrophysics Data System Bibliographic Services. 
Observations reported here were obtained at the MMT Observatory, a joint facility of the University of Arizona and the Smithsonian Institution.
Funding for SDSS-III has been provided by the Alfred P. Sloan Foundation, the Participating Institutions,
 the National Science Foundation, and the U.S. Department of Energy Office of Science. 
The SDSS-III website is http://www.sdss3.org/. 
SDSS-III is managed by the Astrophysical Research Consortium for the Participating Institutions of the SDSS-III Collaboration, 
 including the University of Arizona, the Brazilian Participation Group, Brookhaven National Laboratory, University of Cambridge, 
 Carnegie Mellon University, University of Florida, the French Participation Group, the German Participation Group, Harvard University, 
 the Instituto de Astrofisica de Canarias, the Michigan State/Notre Dame/JINA Participation Group, Johns Hopkins University,
  Lawrence Berkeley National Laboratory, Max Planck Institute for Astrophysics, Max Planck Institute for Extraterrestrial Physics, 
  New Mexico State University, New York University, The Ohio State University, The Pennsylvania State University, University of Portsmouth, 
  Princeton Uni- versity, the Spanish Participation Group, University of Tokyo,
  University of Utah, Vanderbilt University, University of Virginia, University of Washington, and Yale University.

\facility{MMT}

{}

\end{document}